\documentclass[twocolumn,letter]{jpsj2}

\title{
	Tunneling Properties at the Interface 
	between Superconducting 
	Sr$_2$RuO$_4$  \\ and a Ru Microinclusion
}

\author{
	Minoru \textsc{Kawamura}$^{1,2}$, 
	Hiroshi \textsc{Yaguchi}$^{3}$, 
	Naoki \textsc{Kikugawa}$^{3}$,
	Yoshiteru \textsc{Maeno}$^{3,4}$
	and Hideaki \textsc{Takayanagi}$^{1}$
}

\inst{
	$^{1}$NTT Basic Research Laboratories,
		NTT Corporation, Atsugi, Kanagawa 243-0198, Japan \\
	$^{2}$RIKEN, Wako, Saitama 351-0198, Japan \\
	$^{3}$Department of Physics, Graduate School of Science,
		Kyoto University, Kyoto 606-8502, Japan \\
	$^{4}$Kyoto University International Innovation Center,
		Kyoto 606-8501, Japan
}

\abst{
We have investigated
the magnetic field and temperature  dependence 
of the tunneling spectra of the eutectic system Sr$_2$RuO$_4$-Ru.
Electric contacts 
to individual Ru lamellae embedded in Sr$_2$RuO$_4$
enable the tunneling spectra at the interface
between ruthenate and a Ru microinclusion
to be measured.
A zero bias conductance peak (ZBCP) was observed
in the bias voltage dependence of the differential conductance,
suggesting that Andreev bound states are present at the interface.
The ZBCP starts to appear at a temperature well below 
the superconducting transition temperature.
The onset magnetic field of the ZBCP is  also considerably smaller
than the upper critical field
when the magnetic field is parallel to the {\it ab}-plane.
We propose that the difference between the onset of the ZBCP
and the onset of superconductivity can be understood
in terms of the existence of the single-component state
predicted  by Sigrist and Monien. 
}

\kword{Sr$_2$RuO$_4$, $p$-wave superconductor, eutectic, 
	zero bias conductance peak}

\begin{document}
\maketitle


Recent experimental and theoretical studies have revealed that 
Sr$_2$RuO$_4$ is a spin-triplet superconductor\cite{maeno94,mackenzie03}.
An NMR measurement confirmed this first:
the Knight shift of $^{17}$O is not affected
by the superconducting (SC) transition,
indicating that the spin state of the Cooper pair
is triplet\cite{ishida98}.
It was demonstrated by a muon spin relaxation measurement\cite{luke98}
that spontaneous magnetic moments accompany
the SC transition.
This result indicates that 
time reversal symmetry is broken in the SC phase,
suggesting that the orbital part of the order parameter 
of the SC state 
has two components
with a relative phase of $\pi/2$: $k_x + ik_y$.

Amongst the many interesting SC properties of Sr$_2$RuO$_4$,
the enhancement of the SC transition temperature
in the eutectic system\cite{maeno98} is rather interesting.
Under particular growth conditions, 
a single crystal of Sr$_2$RuO$_4$ with Ru lamellae included
is obtained. 
This eutectic system, Sr$_2$RuO$_4$-Ru, shows 
a broad SC transition with an onset of approximately 3 K\cite{maeno98},
called  the 3-K phase.
The manifestation of 3-K phase superconductivity
has been confirmed by ac susceptibility measurements\cite{maeno98} and 
resistivity measurements\cite{maeno98, ando99}. 
These experimental studies suggest that
3-K phase superconductivity is inhomogeneous and filamentary,
and thus it is inferred to occur at the interface between Sr$_2$RuO$_4$ and Ru.

\begin{figure}[t]
	\begin{center}
		\includegraphics[width=8cm]{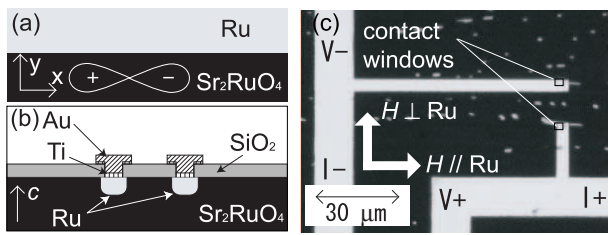}
	\end{center}
	\caption{
		\label{sample}
	(a) Schematic of the interface between Sr$_2$RuO$_4$ and Ru
		modeled by Sigrist and Monien.
	(b) Cross-sectional sketch of the sample.
		Sample structure is described in detail in text.
	(c) Top view of the sample used.
		The {\it c}-axis direction is 
		normal to the picture plane.
		The contact windows are seen at the edge of Ti/Au electrodes.
		Magnetic field directions with respect to the Ru lamella
		are indicated by arrows.
	}
\end{figure}

The anisotropy of the upper critical field
indicates that the Sr$_2$RuO$_4$ side of the interface
essentially sustains the superconductivity\cite{maeno98}.
The temperature dependence of the upper critical magnetic field $H_{\rm c2}$
shows an unusual upturn at low temperatures
when the magnetic field is applied parallel 
to the {\it c}-axis\cite{yaguchi03}.
The low-temperature enhancement
of  $H_{\rm c2}$ for  $H \parallel c$ can be explained
in terms of the coupling of the two components of the order parameter
discussed in Sigrist and Monien's theory\cite{sigrist01}.
The good agreement between  theory and  experiment supports
the description of the 3-K phase by a two-component order parameter
with a relative phase of $\pi$/2.
According to the theory by Sigrist and Monien, 
a single-component order parameter $k_x$ with 
its lobes parallel to the interface shown in Fig.~\ref{sample}(a),
nucleates at the onset of the 3-K phase.
With decreasing temperature and/or magnetic field,
the superconductor undergoes another transition
to the two-component state $k_x + i\varepsilon k_y$
($0 < \varepsilon < 1$)
with broken time reversal symmetry, similar to 
pure Sr$_2$RuO$_4$ without Ru inclusions.

Tunneling spectroscopy has played a crucial role 
in determining the phase structure of {\it d}-wave superconductors\cite{hu94, tanaka95, kashiwaya95},
and thus will be a promising technique for
examination of the transition to the two-component state in the 3-K phase.
Several tunneling spectroscopy measurements have already been performed 
on Sr$_2$RuO$_4$\cite{laube00,upward02}.
A tunneling measurement using cleaved junctions 
of the eutectic system Sr$_2$RuO$_4$-Ru was demonstrated
by Mao {\it et al}\cite{mao01}.
Their results gave  strong evidence that
unconventional superconductivity occurs in the 3-K phase.
They observed a zero bias conductance peak (ZBCP) 
in the voltage dependence of the differential conductance.
The ZBCP is interpreted in terms of an Andreev bound state (ABS)
at the interface between Sr$_2$RuO$_4$ and Ru.

In this paper, 
our experimental study of the ZBCP in the 3-K phase is reported.
A microfabrication technique enabled
the measurement of the differential conductance
at the interface between a single Ru microinclusion and Sr$_2$RuO$_4$.
The ZBCP was observed 
in the voltage dependence of the differential conductance.
We have found that
the ZBCP starts to appear
at a lower temperature and/or magnetic field
than the onset of 3-K phase superconductivity.
The difference between the onset of the ZBCP 
and the onset of 3-K phase superconductivity is discussed
with the help of Sigrist and Monien's theory\cite{sigrist01}.


The eutectic samples of Sr$_2$RuO$_4$-Ru
used in the present experiment were grown
by a floating zone method.
Ru lamellae with typical dimensions of
approximately 1 $\mu$m $\times$ 5 $\mu$m $\times$ 20 $\mu$m
were included in the samples.
Details of the crystal growth and the properties of the eutectic
samples are described in Ref.~\citen{maeno98}.
In order to measure the conductance at the interface
between a single Ru microinclusion and Sr$_2$RuO$_4$,
we fabricated the device illustrated in Fig.~\ref{sample}(b).
The device has pairs of Ti/Au electrodes 
directly attached to individual Ru microinclusions
through 2 $\mu$m $\times$ 3 $\mu$m contact windows.
Figure~\ref{sample}(c) is a top view of the device.
This device was fabricated 
by employing the following microfabrication technique.
First, a 200-nm-thick SiO$_2$ insulating film
was deposited on the polished $ab$-plane surface of the
Sr$_2$RuO$_4$-Ru eutectic by rf sputtering.
Contact windows of 2 $\mu$m $\times$ 3 $\mu$m were etched 
through the SiO$_2$ film by using standard electron beam lithography
and an Ar ion milling technique, such that
the contact windows were placed immediately above the Ru lamellae.
The positions of the Ru lamellae were accurately measured in advance
by an optical microscope which has a sample stage 
equipped with a laser interferometer.
Finally, electrodes of Ti/Au were patterned by means of
electron beam lithography and a lift-off technique.
The thicknesses of the Ti and Au films were
about 20 nm and 80 nm, respectively.

As seen in Fig.~\ref{sample}(c), 
the contact windows are slightly larger than 
the Ru lamellae.
However, a non-superconducting surface layer
with a relatively high resistivity exists 
at the polished surface of the Sr$_2$RuO$_4$ crystal.
It is reported that
the resistance between Sr$_2$RuO$_4$ and
an In wire with a cross-sectional area of 0.05 mm$^2$
ranges from 0.1 to 100 $\Omega$\cite{jin00}.
This leads to the estimated
resistance through the  (2-$\mu$m $\times$ 3-$\mu$m)-sized
contact window to be between 1 k$\Omega$ and 1 M$\Omega$.
The contact resistance between the electrodes
and the Ru metal lamella is expected to be much smaller 
than that between the electrodes and the non-superconducting surface layer.
Therefore, the measurement current will dominantly
flow through the Ru metal lamella.

The differential conductance between two Ru lamellae 
was measured by a two-terminal method as in Fig.\ref{sample}(c).
A lock-in technique was employed with alternating
currents of 10 $\mu$A at 184 Hz.
Temperature was decreased down to 0.3 K
by means of a $^3$He refrigerator.
A magnetic field of up to 4 T was applied
by means of a superconducting solenoid.
A single-axis sample rotator enabled
the magnetic field  to be aligned parallel to the {\it c}-axis
or  the {\it ab}-plane, within a few degrees.

\begin{figure}[t]
	\begin{center}
		\includegraphics[width=7.4cm]{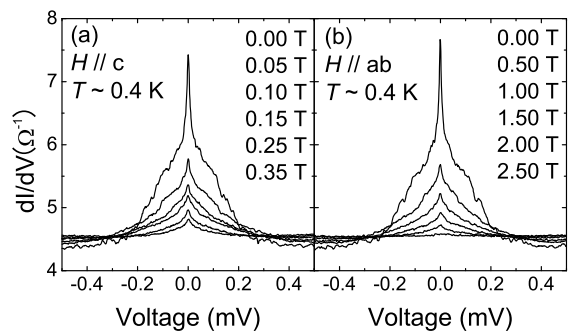}
	\end{center}
	\caption{
		\label{diffres}
		ZBCP observed in the bias voltage dependence of
		${\rm d}I/{\rm d}V$ in sample A  at approximately 0.4 K.
		Magnetic field direction is parallel 
		to the {\it c}-axis in (a)
		and parallel to the {\it ab}-plane and the Ru-plane in (b).
	}
\end{figure}


We measured the tunneling spectra on three junction samples
with different conductances
on the same crystal.
The conductances of the three samples A, B and C
were 7.5, 36 and 0.28 $\Omega^{-1}$, 
respectively.
Tunneling spectra with a sharp conductance peak
at zero bias voltage were obtained for sample A as shown in Fig.~\ref{diffres};
the bias voltage dependence of
the differential conductance ${\rm d}I/{\rm d}V$ at approximately 0.4 K
is plotted for different magnetic fields.
The magnetic field direction is parallel to the {\it c}-axis
in (a)
and parallel to the {\it ab}-plane
and the Ru lamella plane in (b).
The field direction with respect to the Ru lamella plane is
illustrated in Fig.~\ref{sample}(c).
The ZBCP persists to well above the upper critical field
of the 1.5-K phase and thus involves the 3-K phase.

Due to the Joule heating by the bias voltage,
the temperature increased during the measurement.
The increase of temperature was no greater than 0.1 K at 0.5 mV. 
The effect of the Joule heating was negligible 
at the voltages below 0.1 mV.
Because the width of the ZBCP is smaller than 0.2 mV, 
we believe that the Joule heating
affected the feature of the ZBCP negligibly.

The presence of the ZBCP in the spectra
ensures that an ABS is formed at the interface involving the Ru lamellae.
In our measurement configuration,
the resistances of Sr$_2$RuO$_4$, Ru and Ti/Au electrodes
are measured in series.
The resistance of Sr$_2$RuO$_4$ estimated from its resistivity
$\rho_{ab} \approx 1$ $\mu\Omega$cm and
$\rho_c \approx 30$ $\mu\Omega$cm
is much smaller than the measured resistance.
The measured resistance is probably dominated
by the resistance at the Sr$_2$RuO$_4$/Ru interface. 
ZBCP is absent in samples B and C.
Although the magnetic field dependence of the conductance
shows a small peak in sample B,
it can be suppressed by the application 
of a much smaller field (0.2 T) than  $H_{\rm c2}$ of the 3-K phase.
Therefore, the peak is considered to have an origin other than the 3-K phase.
Sample C shows a conductance dip rather than a peak 
in the voltage dependence of the differential conductance.
This feature is characteristic of the Andreev reflection 
at an interface with a large tunnel barrier.
This  suggests that the condition 
of the interface may differ from place to place
within a single piece of crystal.
Further systematic experimental studies
are awaited to allow the condition for the ZBCP to be identified.

\begin{figure}[t]
	\begin{center}
		\includegraphics[width=6.5cm]{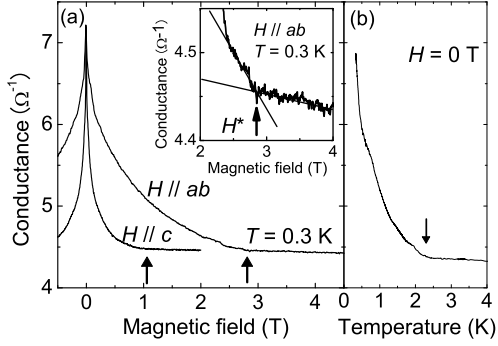}
	\end{center}
	\caption{
		\label{magres}
	(a) Magnetic field dependence of the zero bias conductance
	at $T$ = 0.3 K for $H \parallel c$ (lower curve) 
	and $H \parallel ab \parallel$ Ru-plane (upper curve).
	The onset of the ZBCP is indicated by arrows.
	The inset shows the definition of the onset magnetic field
	of the ZBCP $H^*$.
	(b) Temperature dependence of the zero bias conductance
	in zero magnetic field.	
	The onset of the ZBCP is indicated by an arrow.
	}
\end{figure}

Tunneling spectra on sample A are similar to those
in the cleaved junctions in Ref.~\citen{mao01}.
Sharp conductance peaks together with bell-shaped broad conductance peaks
are seen in zero magnetic field in both experiments.
However, there are still quantitative differences between
the results reported in Ref.~\citen{mao01} 
and those for the present experiment. 
The width of the conductance peak in our experiment
is about half that reported in Ref.~\citen{mao01}.
The quantitative difference between the two experiments 
might be attributed to the difference in the sample structure.
In our sample, the conductance 
associated with two Sr$_2$RuO$_4$/Ru interfaces in series was measured.
In the cleaved junctions in Ref.~\citen{mao01},
many Sr$_2$RuO$_4$/Ru interfaces at the cleavage
were probably measured in parallel.
The size and the number of junctions involved cannot be specified
in the cleaved junction experiment,
making it  difficult to compare the two experiments quantitatively.

\begin{figure}[t]
	\begin{center} 
		\includegraphics[width=5.3cm]{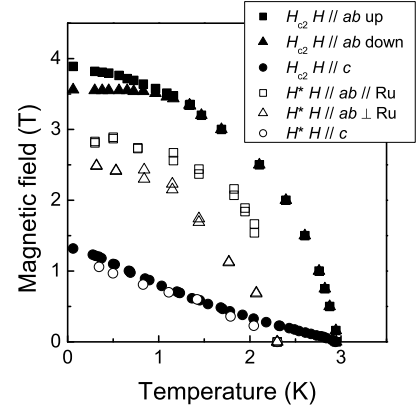}
	\end{center}
	\caption{
		\label{phasediagram}
		Temperature dependence of the 
		onset of the ZBCP $H^*$ and the upper critical magnetic field
		$H_{\rm c2}$.
		Closed symbols represent $H_{\rm c2}$
		and  open symbols represent $H^*$.
	}
\end{figure}

The magnetic field dependence of  
the ZBCP shows distinctly different behavior
for different field directions.
Figure~\ref{magres}(a) shows the magnetic field dependence of the conductance
at $T$ = 0.3 K for $H \parallel c$ and $H \parallel ab \parallel$ Ru-plane.
Here we pay attention to the magnetic field $H^*$
at which the conductance starts to increase with decreasing 
magnetic field.
$H^*$ is defined as the intersection of two tangential lines
as shown in the inset of Fig.~\ref{magres}(a).
It should be noted that $H^*$ corresponds to the 
onset magnetic field of the ZBCP,
which is consistent with the spectra in Fig.~\ref{diffres}.
We compare the temperature dependence of $H^*$
with that of $H_{\rm c2}$.
The temperature dependence of $H^*$ is plotted
in Fig.~\ref{phasediagram} with open symbols
together with the temperature dependence of $H_{\rm c2}$.
The  data for $H_{\rm c2}$ is taken from Ref.\citen{yaguchi03b}.
Similar to the definition of $H^*$,
$H_{\rm c2}$ in Ref.\citen{yaguchi03b}
is defined as the onset of the resistance drop
in a bulk sample from a different batch.
On the other hand, $H_{\rm c2}$ in Ref.\citen{yaguchi03}
is defined as the inflection point 
associated with the resistance drop.
There are remarkable differences between $H^*$ and $H_{\rm c2}$.
The temperature dependence of the upper critical field $H_{\rm c2}$ shows
hysteresis at low temperatures 
only when  magnetic field is applied 
parallel to the {\it ab}-plane\cite{ando99, yaguchi03}.
However, 
no clear hysteresis can be seen in the temperature dependence of $H^*$.
Despite the difference of the sample batch,
$H^*(T)$ and $H_{\rm c2}(T)$ show
a good coincidence for $H \parallel c$
except  at very low fields.
On the other hand, $H^*(T)$ is considerably smaller than 
$H_{\rm c2}(T)$ for $H \parallel ab$ and $H \approx 0$ T.
We suggest that the ZBCPs are absent near the onset of the 3-K phase
in a magnetic field $H \parallel ab$ and $H \approx 0$.
While the onset temperature of the ZBCP in the zero magnetic field
is 2.3 K as shown in Fig.~\ref{magres}(b),
a standard resistance measurement on a sample
from the same batch has yielded $T_{\rm c}$ = 2.7 K (onset),
higher than the onset of the ZBCP.
This also supports the above conclusion.

Superconducting properties of the 3-K phase
such as $T_{\rm c}$ and $H_{\rm c2}$
may differ from place to place, which would cause 
a discrepancy between $H^*$ and $H_{\rm c2}$.
However, there are reasons why $H^*$ and $H_{\rm c2}$
are different in origin, as suggested in Fig.\ref{phasediagram}.
The phase diagrams in Refs.\citen{ando99} and \citen{yaguchi03}
are very similar and $H_{\rm c2}$ exhibits hysteresis
for $H \parallel ab$, unlike $H^*$.
In fact, it would be desirable to determine
both $H_{\rm c2}$ and $H^*$ in the same junction.
However, the resistance measured was dominated by the 
interface resistance, so that the resistance change
due to the SC transition was barely detected.

We propose that the difference between $H^*$ and $H_{\rm c2}$
can be attributed to the phase-sensitive nature
of the ABS\cite{tanaka95, kashiwaya95}.
When Andreev reflection occurs at a superconductor/normal-metal interface, 
the reflected quasiparticle gains the phase of the pair potential.
Because an ABS is a consequence 
of the interference of the injected and reflected quasiparticles, 
the condition for the formation of the ABS strongly depends
on the phase structure of the pair potential.
According to a theory for a normal metal/anisotropic superconductor
junction\cite{tanaka95},
the ABS is not formed for any incident angle 
in the case of a single-component order parameter $k_x$
with its lobes parallel to the interface,
as depicted in Fig.~\ref{sample}(a).
On the other hand, the other component $k_y$
with its lobes perpendicular to the interface gives rise to the ZBCP.
The ZBCP is also theoretically predicted 
in the case of the chiral order parameter $k_x + i\varepsilon k_y$ 
\cite{honerkamp98, tanaka02}.
Thus the appearance of the ZBCP is dependent on the 
phase structure of the pair potential.
The absence of the ZBCP near the onset of the 3-K phase 
for $H = 0$ T and $H \parallel ab$ suggests that
a SC state which does not give rise to the ZBCP
nucleates at the onset of the 3-K phase.

This interpretation of the absence of the ZBCP
near the onset of the 3-K phase for $H = 0$ T and $H \parallel ab$
is supported by the theoretical predictions
by Sigrist and Monien\cite{sigrist01}.
The theory predicts
a nucleation of a single-component order parameter with
time reversal symmetry conserved.
Due to the lower symmetry at the interface,
the single-component order parameter with its lobes
parallel to the interface $k_x$,
which is illustrated in Fig.~\ref{sample}(a),
is favored at the onset of the 3-K phase.
Further cooling gives rise to the nucleation of the other component
$k_y$ with a relative phase of $\pi/2$,
leading to the chiral state $k_x + i\varepsilon k_y$.
Because the single-component order parameter $k_x$ 
does not give rise to the ZBCP,
the absence of the ZBCP at the onset of the 3-K phase
for $H = 0$ T and $H \parallel ab$  can be attributed
to the presence of the single-component order parameter $k_x$.

The theory also predicts 
the coupling of the two components
of the order parameter
in the presence of magnetic field $H$ $\parallel c$.
As the two-component state (chiral state)
represents a state with a finite orbital angular momentum
along the {\it c}-axis, 
a magnetic field component parallel to the {\it c}-axis
causes the two-component state to be stabilized,
unlike $H = 0$ T or $H \parallel ab$.
The coupling of the two components is 
strengthened with decreasing temperature and 
increasing the magnetic field component parallel to the {\it c}-axis.
The scenario of the coupling of the two components
is supported by recent experimental results\cite{yaguchi03}.
It is possible to consider that
the coupling of the two components causes 
the ZBCP to emerge at the onset of the 3-K phase.
The good coincidence of $H^*$ and $H_{\rm c2}$ for $H \parallel c$
can be attributed to the coupling of the two components.

The relation between $H^*$ and $H_{c2}$ shows
a distinct difference, depending on the magnetic field direction
as discussed above.
It is further shown that $H^*$ exhibits in-plane anisotropy
when the magnetic field is rotated in the {\it ab}-plane.
The electric contacts to the individual Ru lamellae enabled
the in-plane anisotropy of $H^*$ to be observed.
In Fig.~\ref{phasediagram}, the temperature dependence of $H^*$
for the magnetic field parallel and perpendicular to the Ru lamella plane
is plotted with open squares and open triangles, respectively.
$H^*$ for $H \perp$ Ru-plane is 89\% 
of that for  $H \parallel$ Ru-plane at $T$ = 0.3 K.
On the other hand, no significant dependence of $H_{\rm c2}$
on the field direction relative to the crystallographic axes
has been observed in a resistivity measurement
in the eutectic system\cite{yaguchipc}.
This is probably because
many Ru lamellae oriented in different directions are
involved in the resistivity measurement, and thus
the anisotropy around each Ru inclusion will be averaged out.
Because the 3-K phase is inferred to be surface superconductivity
around  Ru lamellae\cite{maeno98,ando99},
the critical magnetic field for  $H \perp$ Ru-plane is
expected to be smaller than that for  $H \parallel$ Ru-plane,
as was discussed by Matsumoto {\it et al}.\cite{matsumoto03}
The observed in-plane anisotropy of $H^*$ is considered 
to reflect the anisotropy of $H_{\rm c2}$ with respect
to the direction of the Ru lamella.
The in-plane anisotropy of $H^*$ relative to the Ru lamella
directly supports the proposition that 3-K phase superconductivity
is filamentary and occurs at surfaces surrounding Ru\cite{maeno98, ando99}.

To summarize, we have developed a device 
for the measurement of the conductance at the interface
between  Sr$_2$RuO$_4$ and a single Ru microinclusion.
The tunneling spectra with the ZBCP were obtained 
in the voltage dependence of the differential conductance.
The magnetic field and temperature dependence 
of the onset of the ZBCP is discussed in comparison with
the onset of the superconductivity $T_{\rm c}$ and $H_{\rm c2}$.
We propose that the difference between 
the onset of the ZBCP and the onset of the 3-K phase
can be attributed to the presence of a superconducting state
which is not accompanied by the ABS.
This is consistent with the theoretical prediction of
the nucleation of the single-component state $k_x$,
which does not give rise to the ABS, at the onset of the 3-K phase.
We have also observed in-plane anisotropy of $H^*$
with respect to the direction of Ru lamellae.

Fruitful discussions with
M. Sigrist, M. Matsumoto and Y. Tanaka are gratefully acknowledged.
We acknowledge the technical support of 
T. Meno, T. Akazaki, J. Nitta and T. Kimura.
This work has been 
supported in part by a Grant-in-Aid for
Scientific Research 
from the Japan Society for the Promotion of Science.

\end{document}